\documentstyle[IJFCS]{article}

\begin{document}
\copyrightheading

\symbolfootnote

\textlineskip

\begin{center}

\fcstitle{SEPARATION BETWEEN CLASSICAL\\
          AND QUANTUM WINNING STRATEGIES\\
          FOR THE MATCHING GAME}

\vspace{24pt}

{\authorfont IVAN FIAL\'IK}

\vspace{2pt}

\smalllineskip
{\addressfont Faculty of Informatics, Masaryk University\\
        Botanick\'a 68a, 602 00 Brno, Czech Republic}

\vspace{20pt}
\publisher{(received date)}{(revised date)}{Editor's name}

\end{center}

\alphfootnote

\begin{abstract}
Communication complexity is an area of classical computer science which studies how much communication is necessary to solve various distributed computational problems. Quantum information processing can be used to reduce the amount of communication required to carry out some distributed problems. We speak of pseudo-telepathy when it is able to completely eliminate the need for communication. The matching game is the newest member of the family of pseudo-telepathy games. After introducing a general model for pseudo-telepathy games, we focus on the question what the smallest size of inputs is for which the matching game is a pseudo-telepathy game.

\end{abstract}

\textlineskip
\section{Introduction}
Quantum information processing allows us to solve problems that we are not able to solve in the classical world at all or at least that we are not able to solve efficiently. This is true also in the field of communication complexity. The first convincing evidence that quantum communication protocols can be more efficient than classical ones was given in 1998 by Buhrman, Cleve and Wigderson \cite{qvccac}. They found a problem whose quantum communication complexity is exponentially better than classical communication complexity in the error-free model. One year later, Raz proposed a problem for which this exponential separation holds also in the bounded-error model \cite{esoqaccc}. Since quantum entanglement provides us with strong non-local correlations, one can ask whether it can be used even to completely eliminate the need for communication. Of course, we are interested only in such problems for which this does not hold in the classical world. On one hand, the answer is negative if we consider the standard communication complexity model \cite{scqrtdc} in which parties compute a value of some function on their inputs and the whole result of the computation must become known to at least one party. Otherwise, faster-than-light communication would be possible which would contradict the Relativity Theory. On the other hand, if each party has its own input, computes its own output and we are interested only in non-local correlations between the inputs and the outputs, then the answer is positive. Such problems are often described using a terminology of the game theory and they are usually called pseudo-telepathy games.

Apart from the fact that they can be seen as distributed problems which can be solved without any form of direct communication between the parties, there is one more reason to be interested in pseudo-telepathy games. They offer an alternative way to show that the physical world is not \emph{local realistic}, the result which is usually proved using some form of the Bell inequality \cite{oteprp}. \emph{Locality} means that no action performed at a location $A$ can have an instantenous (faster than light) effect at a remote location $B$. \emph{Realism} means that every characteristic about the physical system that can be measured is already determined before the actual measurement. Therefore, we can say that it exists independently of the measurement. Unfortunately, the Bell inequality is not very easy to explain because it involves nontrivial probabilistic arguments. It would be very convenient if we could demonstrate an observable behaviour which is obviously impossible in the classical world. Pseudo-telepathy games are of interest because some of them are very simple and one can explain that there is no classical winning strategy for them in several minutes almost to anyone.  

In order to be able to describe what a pseudo-telepathy game is, we explain at first what we mean by the term two party game. A \emph{two party game} $G$ is a sextuple $(X, Y, A, B, P, W)$ where $X, Y$ are \emph{input sets}, $A, B$ are \emph{output sets}, $P$ is a subset of $X \times Y$ known as a \emph{promise} and $W \subseteq X \times Y \times A \times B$ is a relation among the input sets and the output sets which is called a \emph{winning condition}. Before the game begins, the parties, usually called Alice and Bob, are allowed to discuss strategy and exchange any amount of classical information, including values of random variables. They may also share an unlimited amount of quantum entanglement. Afterwards, Alice and Bob are separated from each other and they are not able to communicate any more till the end of the game. In one \emph{round of the game}, Alice is given an input $x \in X$ and she is required to produce an output $a \in A$. Similarly, Bob is given an input $y \in Y$ and he is required to produce an output $b \in B$. The pairs $(x, y)$ and $(a, b)$ are called a \emph{question} and an \emph{answer}, respectively. We say that Alice and Bob \emph{win the round} if either $(x, y) \notin P$ or $(x, y, a, b) \in W$. Alice and Bob \emph{win the game} if they have won all the rounds of it. A \emph{strategy} of Alice and Bob is said to be \emph{winning} if it always allows them to win. 

We say that a two-party game is \emph{pseudo-telepathic} if there is no classical winning strategy, but there is a winning strategy, provided Alice and Bob share entanglement. The origin of this term can be explained in the following way. Suppose that scientists who know nothing about quantum computing witness Alice and Bob playing some pseudo-telepathy game. More precisely, suppose that the players are very far from each other, they are given their inputs at the same time and have to produce their outputs in time shorter than time required by light to trave1 between them. If Alice and Bob answer correctly in a sufficiently long sequence of rounds, the scientists will conclude that Alice and Bob can communicate somehow. But according to classical physics, communication between the players is impossible. Therefore, the scientists will be made to believe that Alice and Bob are able to communicate in the way unknown to classical physics. Now, one of possible explanations will be that the players are endowed with telepathic powers. A survey of pseudo-telepathy games can be found in \cite{qp}. The definition of these games can be easily generalized to more than two players.

A classical strategy $s$ for a pseudo-telepathy game $G$ is \emph{deterministic} if there are functions $s_{A} : X \to A$ and $s_{B}: Y \to B$ such that for each question $(x, y) \in X \times Y$, the only possible answer of Alice and Bob is the pair $(s_{A}(x), s_{B}(y))$. The \emph{success} $\omega_{s}(G)$ {of a deterministic strategy} $s$ is defined as the proportion of questions from the promise $P$ for which $s$ produces a correct answer. Clearly, this number can by interpreted as the probability that the strategy $s$ succeeds on a given question which is chosen uniformly and randomly. We denote with $\omega_{d}(G)$ the maximal success of a deterministic strategy for the game $G$: 
\begin{equation}
\omega_{d}(G) = \max_{s} \frac{\{(x, y) \in P \ | \ (x, y, s_{A}(x), s_{B}(y)) \in W\}}{|P|}.
\end{equation}

Alice and Bob can also use a classical randomized strategy for $G$. Any randomized strategy can be seen as a probability distribution over a finite set of deterministic strategies. Therefore, if questions are chosen uniformly and randomly, the probability of winning the game $G$ using a randomized strategy cannot be greater than $\omega_{d}(G)$ \cite{qp}.
 
This paper examines how successful classical players can be at the matching game. This game is described in the next section. Classical winning strategies for inputs of size 4 and for inputs of size 6 are proposed in Section 3. In Section 4, we show that there is no classical winning strategy if the input size is greater than 6.

\section{The Matching Game}
The matching game is the youngest member of the family of pseudo-telepathy games. It was proposed by Buhrman and Kerenidis in 2004 \cite{qpg}.

\begin{definition}
\item A perfect matching $M$ on the set $\{0, \ldots, m - 1\}$, where $m$ is even, is a partition of this set into $\frac{m}{2}$ sets, each of cardinality 2. We define $M_{m}$ as the set of all perfect matchings on $\{0, \ldots, m - 1\}$.
\end{definition}
\subsection{The game}
Alice receives a bit string $x = x_{0}x_{1}\cdots x_{m - 1}$ and Bob receives a perfect matching $y \in M_{m}$. The task for Alice is to output a string $a \in \{0, 1\}^{\lceil \log m \rceil}$. The task for Bob is to output a set $\{b_{1_{1}}, b_{1_{2}}\} \in y$ and a string $b_{2} \in \{0, 1\}^{\lceil \log m \rceil}$. The players win the round if and only if 
\begin{equation}
x_{b_{1_{1}}} \oplus x_{b_{1_{2}}} = (\bar{b}_{1_{1}} \oplus \bar{b}_{1_{2}}) \cdot (a \oplus b_{2})
\label{eq.1}
\end{equation}
where $u \cdot v = \bigoplus_{i = 1}^{n}(u_{i} \wedge v_{i})$ and $\bar{b}_{1_{1}},  \bar{b}_{1_{2}} \in \{0, 1\}^{\lceil \log m \rceil}$. The exclusive-or operator is applied on bits on the left side of the equation and is applied bit-wise on bit strings on the right side. The bit string $\bar{b}_{1_{1}}$ is a binary representation of the number $b_{1_{1}}$ in which the most significant bit of $b_{1_{1}}$ is preceded by $k_{1_{1}}$ zero bits where $k_{1_{1}} = \lceil \log m \rceil - \lfloor \log b_{1_{1}} \rfloor - 1$. Similarly, the bit string $\bar{b}_{1_{2}}$ is a binary representation of the number $b_{1_{2}}$. 

A formal definition of the matching game is given in Table \ref{tab.1}.
\begin{table}[htbp]
\tcaption{The matching game.}
\begin{center}
\begin{tabular}{|c|c|}
   \hline
   $X$ & $\{0, 1\}^{m}$ where m is even      \\
   \hline
   $Y$ & $M_{m}$                             \\
   \hline
   $A$ & $\{0, 1\}^{\lceil \log m \rceil}$  \\
   \hline
   $B$ & $\{\{b_{1_{1}}, b_{1_{2}}\} \ | \ b_{1_{1}}, b_{1_{2}} \in \{0, 1, \ldots, m - 1\}\} \times \{0, 1\}^{\lceil \log m \rceil}$     \\
   \hline
   $P$ & $X \times Y$                        \\
   \hline
   $W$ & $x_{b_{1_{1}}} \oplus x_{b_{1_{2}}} = (\bar{b}_{1_{1}} \oplus \bar{b}_{1_{2}}) \cdot (a \oplus b_{2}) \wedge \{b_{1_{1}}, b_{1_{2}}\} \in y$   \\
   \hline   
\end{tabular}
\label{tab.1}
\end{center}
\end{table}

A quantum winning strategy for the matching game and also the proof that it always succeeds can be found in \cite{qpg}. The proof of the non-existence of a  classical winning strategy for the matching game is based on the exponential separation between quantum and classical one-way communication complexity of the hidden matching problem \cite{esoqacocc, qpg}.

\section{Classical Winning Strategies for $m = 4$ and $m = 6$}
The above asymptotic result tells us only that for large enough inputs, there is no classical winning strategy for the matching game. But to be able to perform practical experiments, it is important to know exactly the smallest size of inputs with this property. Obviously, there is a classical winning strategy for $m = 2$ because there is only one perfect matching on the set $\{0, 1\}$. We propose classical winning strategies both for $m = 4$ and $m = 6$. These strategies are both obtained as a straightforward consequence of the following lemma which tells us that for each input size, there is a classical strategy which is winning if we properly restrict the set of questions Alice and Bob can be given.

\begin{definition}
For a positive integer $m$, we denote with $W_{m}$ the set $\{0\} \cup \{2^{i} \ | \ i \in \{0, 1, \ldots, \lceil \log m \rceil - 1\}\}$.
\end{definition}

\begin{lemma}
Let $m > 0$ be an even integer. Suppose that Alice is given an input $x = x_{0}x_{1}\cdots x_{m - 1}$ and that Bob's input $y$ contains a pair $\{w_{1}, w_{2}\} \subset W_{m}$. If Alice outputs the string $a = (x_{0} \oplus x_{2^{\lceil \log m \rceil - 1}}) (x_{0} \oplus x_{2^{\lceil \log m \rceil - 2}}) \cdots (x_{0} \oplus x_{1})$ and Bob outputs the pair $b = (\{w_{1}, w_{2}\}, 0^{\lceil \log m \rceil})$, the players will win.
\label{lem.1}
\end{lemma}
\proof{
Let $str^{m}(i, j)$, where $i, j \in \{0, \ldots, \lceil \log m \rceil - 1\}$, be the bit string of length $\lceil \log m \rceil$ such that
\begin{eqnarray*}
\begin{array}{ll}
str^{m}(i, j)_{k} = 1 \ & \ \textrm{if } k = i \vee k = j\\
str^{m}(i, j)_{k} = 0 \ & \ \textrm{otherwise.}\\
\end{array}
\end{eqnarray*}
We show that for the players' inputs $x$ and $y$, respectively, and their outputs $a$ and $b$, respectively, the equation $$x_{w_{1}} \oplus x_{w_{2}} = (\bar{w}_{1} \oplus \bar{w}_{2}) \cdot (a \oplus 0^{\lceil \log m \rceil})$$ is satisfied. Without loss of generality, four distinct cases are sufficient to consider:
\begin{itemize}
   \item If $x_{0} = 0$, $w_{1} = 0$ and $w_{2} = 2^{j}$, then the right side of the equation can be transformed in the following way: $$str^{m}(j, j) \cdot x_{2^{\lfloor \log m \rfloor - 1}}x_{2^{\lfloor \log m \rfloor - 2}}\cdots x_{1} = x_{w_{2}} = x_{w_{1}} \oplus x_{w_{2}},$$
   \item if $x_{0} = 0$, $w_{1} = 2^{i}$ and $w_{2} = 2^{j}$, then the right side of the equation can be transformed in the following way: $$str^{m}(i, j) \cdot x_{2^{\lfloor \log m \rfloor - 1}}x_{2^{\lfloor \log m \rfloor - 2}}\cdots x_{1} = x_{w_{1}} \oplus x_{w_{2}},$$
   \item if $x_{0} = 1$, $w_{1} = 0$ and $w_{2} = 2^{j}$, then the right side of the equation can be transformed in the following way: $$str^{m}(j, j) \cdot \neg x_{2^{\lfloor \log m \rfloor - 1}}\neg x_{2^{\lfloor \log m \rfloor - 2}}\cdots \neg x_{1} = \neg x_{w_{2}} = x_{w_{1}} \oplus x_{w_{2}},$$
   \item if $x_{0} = 1$, $w_{1} = 2^{i}$ and $w_{2} = 2^{j}$, then the right side of the equation can be transformed in the following way: $$str^{m}(i, j) \cdot \neg x_{2^{\lfloor \log m \rfloor - 1}}\neg x_{2^{\lfloor \log m \rfloor - 2}}\cdots \neg x_{1} = \neg x_{w_{1}} \oplus \neg x_{w_{2}} = x_{w_{1}} \oplus x_{w_{2}}.$$
\end{itemize}
}

\begin{theorem}
There is a classical winning strategy for the matching game for $m = 4$ and also for $m = 6$.
\end{theorem}
\proof{
For $m = 4$, Lemma \ref{lem.1} gives us the following deterministic strategy:

\begin{enumerate}
   \item For an input $x = x_{0}x_{1}x_{2}x_{3}$, Alice outputs a string $a = a_{0}a_{1}$ where $a_{0} = x_{0} \oplus x_{2}$ and $a_{1} = x_{0} \oplus x_{1}$,
   \item for an input $y$, Bob outputs a pair $(\{w_{1}, w_{2}\}, 00)$ where $\{w_{1}, w_{2}\} \subset \{0, 1, 2\}$ and $\{w_{1}, w_{2}\} \in y$.
\end{enumerate}   
This strategy is depicted in Figure \ref{fig.1}.

It follows from Lemma \ref{lem.1} that each deterministic strategy which satisfies simultaneously the following conditions succeeds for all possible inputs of size 6:
\begin{enumerate}
   \item For an input $x = x_{0}x_{1}\cdots x_{5}$, Alice outputs a string $a = a_{0}a_{1}a_{2}$ where $a_{0} = x_{0} \oplus x_{4}$, $a_{1} = x_{0} \oplus x_{2}$ and $a_{2} = x_{0} \oplus x_{1}$,
  \item for an input $y$, Bob outputs a pair $(\{w_{1}, w_{2}\}, 000)$ where $\{w_{1}, w_{2}\} \subset \{0, 1, 2, 4\}$ and $\{w_{1}, w_{2}\} \in y$.
\end{enumerate}  
One possible winning strategy for inputs of size 6 is depicted in Figure 2.

\begin{figure}[htbp]
\caption{Classical winning strategy for $m = 4$.}
\begin{center}
\begin{tabular}{|c|c|}
   \hline
      $x$ & $s_{A}(x)$ \\
   \hline
      0000, 0001, 1110, 1111 & 00 \\
   \hline
      0100, 0101, 1010, 1011 & 01 \\
   \hline
      0010, 0011, 1100, 1101 & 10 \\
   \hline
      1000, 1001, 0110, 0111 & 11 \\
   \hline   
\end{tabular} 
\begin{tabular}{|c|c|}
   \hline
      $y$ & $s_{B}(y)$ \\
   \hline
      \{\{0, 1\}, \{2, 3\}\} & (\{0, 1\}, 00) \\
   \hline
      \{\{0, 2\}, \{1, 3\}\} & (\{0, 2\}, 00) \\
   \hline
      \{\{1, 2\}, \{0, 3\}\} & (\{1, 2\}, 00) \\
   \hline   
\end{tabular}
\\
\label{fig.1}
\end{center}
\end{figure}
\begin{figure}[t]
\caption{Classical winning strategy for $m = 6$.}
\begin{center}
\begin{tabular}{|c|c|}
   \hline
      $x$ & $s_{A}(x)$ \\
   \hline
      000000, 000001, 000100, 000101, &      \\
      111010, 111011, 111110, 111111  & 000  \\
   \hline
      000010, 000011, 000110, 000111, &      \\
      111000, 111001, 111100, 111101  & 100  \\
   \hline
      001000, 001001, 001100, 001101, &      \\
      110010, 110011, 110110, 110111  & 010  \\
   \hline
      001010, 001011, 001110, 001111, &      \\
      110000, 110001, 110100, 110101  & 110  \\
   \hline
      010000, 010001, 010100, 010101, &      \\
      101010, 101011, 101110, 101111  & 001  \\
   \hline
      010010, 010011, 010110, 010111, &      \\
      101000, 101001, 101100, 101101  & 101  \\
   \hline
      011000, 011001, 011100, 011101, &      \\
      100010, 100011, 100110, 100111  & 011  \\
   \hline
      011010, 011011, 011110, 011111, &      \\
      100000, 100001, 100100, 100101  & 111  \\    
   \hline
\end{tabular}
\\
\label{fig.2}
\end{center}
\end{figure}
\begin{figure}[!]
\begin{center}
\begin{tabular}{|c|c|}
   \hline
      $y$ & $s_{B}(y)$ \\
   \hline
      \{\{0, 1\}, \{2, 3\}, \{4, 5\}\}, &                 \\
      \{\{0, 1\}, \{2, 5\}, \{3, 4\}\}  & (\{0, 1\}, 000) \\
   \hline
      \{\{0, 2\}, \{1, 3\}, \{4, 5\}\}, &                 \\
      \{\{0, 2\}, \{1, 5\}, \{3, 4\}\}  & (\{0, 2\}, 000) \\
   \hline
      \{\{0, 4\}, \{1, 2\}, \{3, 5\}\}, &                 \\
      \{\{0, 4\}, \{1, 3\}, \{2, 5\}\}, &                 \\
      \{\{0, 4\}, \{1, 5\}, \{2, 3\}\}  & (\{0, 4\}, 000) \\
   \hline
      \{\{0, 3\}, \{1, 2\}, \{4, 5\}\}, &                 \\
      \{\{0, 5\}, \{1, 2\}, \{3, 4\}\}  & (\{1, 2\}, 000) \\
   \hline
      \{\{0, 2\}, \{1, 4\}, \{3, 5\}\}, &                 \\
      \{\{0, 3\}, \{1, 4\}, \{2, 5\}\}, &                 \\
      \{\{0, 5\}, \{1, 4\}. \{2, 3\}\}  & (\{1, 4\}, 000) \\
   \hline
      \{\{0, 1\}, \{2, 4\}, \{3, 5\}\}, &                 \\
      \{\{0, 3\}, \{1, 5\}, \{2, 4\}\}, &                 \\
      \{\{0, 5\}, \{1, 3\}, \{2, 4\}\}  & (\{2, 4\}, 000) \\
   \hline 
\end{tabular}
\\
\end{center}
\end{figure}
}

\section{Classical Winning Strategies for $m \ge 8$}
This section investigates whether there is a classical winning strategy for the matching game for $m \ge 8$. The task is carried out using some pieces of knowledge from the graph theory. Therefore, we begin this section with several necessary definitions regarding graphs and their properties.

\begin{definition}
A (undirected) graph $G$ is an ordered pair $G = (V, E)$ where $V$ is a set of vertices and $E$ is a set of two-element sets of vertices. These sets are called edges.
\end{definition}

\begin{definition}
Let $G = (V,E)$ be a graph. A path in $G$ is a sequence $v_{0}, v_{1}, \ldots, v_{n}$, where $n$ is a non-negative integer, of mutually different vertices such that for each $i \in \{0, \ldots, n - 1\}$, it holds that $\{v_{i}, v_{i + 1}\} \in E$.
\end{definition}

\begin{definition}
Let $G = (V,E)$ be a graph. The distance $d_{G}(u, v)$ of vertices $u, v \in V$ in $G$ is the smallest number $n$ for which there is a path $v_{0}, v_{1}, \ldots, v_{n}$ in $G$ such that $v_{0} = u$ and $v_{n} = v$.
\end{definition}

\begin{definition}
Let $G = (V, E)$, $G' = (V', E')$ be graphs. We say that $G'$ is a subgraph of $G$ if $V' \subseteq V$ and $E' \subseteq E$. Moreover, $G'$ is said to be an induced subgraph of $G$ if for any vertices $u, v \in V'$, it holds that $\{u, v\} \in E'$ if and only if $\{u, v\} \in E$.
\end{definition}

\begin{definition}
Let $G = (V,E)$ be a graph and let $G' = (V', E')$ be its induced subgraph such that $V' \neq \emptyset$. We say that $G'$ is a connected component (or only component) in $G$ if the following two conditions hold simultaneously:
\begin{enumerate}
   \item There are no vertices $u \in V'$ and $v \in V \setminus V'$ such that $\{u, v\} \in E$,
   \item for any vertices $u, v \in V'$, there is a path $v_{0}, v_{1}, \ldots, v_{n}$ in $G'$ such that $v_{0} = u$ and $v_{n} = v$.
\end{enumerate}
\end{definition}

\begin{definition}
Let $G = (V,E)$ be a graph. We say that $G$ has cardinality (has size) $n$ if $|V| = n$.
\end{definition}   

Now we will proceed in the following way. At first, we assign to each classical deterministic winning strategy $s$ a set of bit strings of length $m$ and a set of subsets of cardinality 2 of the set $\{0, 1, \ldots, m - 1\}$. Then we examine properties of these sets and show that for $m \ge 8$ such sets cannot exist. We conclude that for $m \ge 8$, there is no classical deterministic winning strategy. Since by fixing random variables we can turn any classical randomized winning strategy into a deterministic one, this means that there is no classical winning strategy at all.

\begin{definition}
Let $s$ be any classical deterministic strategy for the matching game for some $m$. We define a graph $G_{s} = (V, E_{s})$ where $V = \{0, 1, \ldots, m - 1\}$ and $E_{s}$ is the set of all elements of the set $W = \{\{i, j\} \ | \ i, j \in \{0, 1, \ldots, m - 1\}\}$ which Bob produces as a part of at least one of his outputs using the strategy $s$.
\label{def.1}
\end{definition}

\begin{lemma}
Let $m > 0$ be an even integer. Suppose that there is a classical deterministic winning strategy $s$ for the matching game for $m$. Then there is a set $R$ of bit strings of length $m$ such that the following conditions hold simultaneously:
\begin{enumerate}
 \item $|R| \ge \frac{2^{m}}{2^{\lceil log \ m \rceil}}$,
 \item the graph $G_{s}$ contains a component of cardinality greater than $\frac{m}{2}$,
 \item for each $\{i, j\} \in E_{s}$, the parity of bits on positions $i$ and $j$ is the same for every $r \in R$.
\end{enumerate}
\label{lem.2} 
\end{lemma} 
\proof{
\begin{enumerate}
  \item There are $2^{m}$ possible inputs and $2^{\lceil log \ m \rceil}$ possible outputs for Alice. Therefore, there are at least $\frac{2^{m}}{2^{\lceil log \ m \rceil}}$ inputs for which Alice produces the same output using $s$. We take as the set $R$ some set of Alice's inputs with this property whose cardinality is at least $\frac{2^{m}}{2^{\lceil log \ m \rceil}}$.
  \item Let us admit that the graph $G_{s}$ does not contain a component of cardinality greater than $\frac{m}{2}$. We show that there is at least one Bob's input for which the strategy $s$ is not defined. In other words, we show that there is a perfect matching $y$ on the set $\{0, 1, \ldots, m - 1\}$ such that for each $\{i, j\} \in y$, $i$ and $j$ are in different components of $G_{s}$. This result provides us with a contradiction because the strategy $s$ is deterministic.

Let $C_{1}, \ldots, C_{k}$ be all the components of the graph $G_{s}$. Suppose without loss of generality that for each $i \in \{1, \ldots, k - 1\}$, the component $C_{i}$ has greater or equal cardinality than the component $C_{i + 1}$. We describe a simple procedure to construct the perfect matching $y$. We begin with $y = \emptyset$. Then we repeat as long as possible the following step. We try to find the greatest index $j \in \{2, \ldots, k\}$ such that the component $C_{j}$ contains a vertex which has not been inserted in $y$ so far. Let us denote with $u_{1}, \ldots, u_{l}$ all the vertices from $C_{j}$ with this property. Since the component $C_{j - 1}$ has greater or equal cardinality than the component $C_{j}$ and we proceed from components of smaller size to components of greater size, there certainly are mutually different vertices $v_{1}, \ldots, v_{l}$ in $C_{j - 1}$ which have not been inserted in $y$ so far. Now for each $i \in \{1, \ldots, l\}$, we insert the set $\{u_{i}, v_{i}\}$ in $y$. If we are not able to find the index $j$, two possible cases can be distinguished. If the component $C_{1}$  does not contain a vertex which has not been inserted in $y$ so far, then there is nothing more to do. On the contrary, if $C_{1}$ contains $2i$ vertices, where $i$ is a non-negative integer, with this property, we remove $i$ sets of vertices from $y$, assign the vertices from $C_{1}$ to vertices from the removed pairs and insert the sets we have obtained in $y$. In both cases we get the perfect matching $y$ which gives us the desired contradiction.   
 
  \item Let $x$, $x'$ be any elements of $R$ and let $\{b_{1_{1}}, b_{1_{2}}\}$ be any element of $E_{s}$. If Bob's input is $y \in Y$ such that $s_{B}(y) = (\{b_{1_{1}}, b_{1_{2}}\}, b_{2})$, for some $b_{2}$, the right side of the equation (\ref{eq.1}) will be the same both for $x$ and $x'$. Since $s$ is a winning strategy, it follows that the parity of bits on positions $b_{1_{1}}$ and $b_{1_{2}}$ is the same both for $x$ and $x'$. Since $x$ and $x'$ has been arbitrarily chosen from $R$, the parity of bits on positions $b_{1_{1}}$ and $b_{1_{2}}$ has to be the same for all elements of $R$. This holds for all pairs of positions from $E_{s}$ because the set $\{b_{1_{1}}, b_{1_{2}}\}$ has been arbitrary as well.
\end{enumerate}
} 

Our goal is to show that for $m \ge 8$, the sets $R$ and $E_{s}$ from Lemma \ref{lem.2} cannot exist. For this purpose, we slightly modify the definition of the graph colouring problem.

\begin{definition}
Let $G = (V, E)$ be a graph and let $h : E \to \{0, 1\}$ be a function. We say that $G$ is colourable according to $h$ if there is a function $c : V \to \{0, 1\}$ such that for each $\{u, v\} \in E$ it holds that $c(u) \oplus c(v) = h(\{u, v\})$. The function $c$ is said to be a colouring of the graph $G$ according to $h$.
\end{definition}

\begin{lemma}
The last condition from Lemma \ref{lem.2} holds for a set $R$ of bit strings of length $m$ and a set $T$ of elements of the set $W = \{\{i, j\} \ | \  i, j \in \{0, 1, \ldots, m - 1\}\}$ if and only if there is a function $h : T \to \{0, 1\}$ for which $|R|$ various colourings of the graph $G = (V, T)$, where $V = \{0, 1, \ldots, m - 1\}$, according to $h$ exist.
\label{lem.3}
\end{lemma}
\proof{
$(\Rightarrow)$ Suppose that for a set $R$ of bit strings of length $m$ and a set $T$ of elements of the set $W$, the last condition from Lemma \ref{lem.2} holds. We intend to find a function $h : T \to \{0, 1\}$ for which $|R|$ various colourings of the graph $G$ according to $h$ exist. Let $r$ be any element of $R$. The function $h$ is defined by $h(\{u, v\}) = r_{u} \oplus r_{v}$, for each $\{u, v\} \in T$. Since every $r \in R$ can be transformed to a colouring $c_{r}$ of the graph $G$ according to $h$ by $c_{r}(u) = r_{u}$, where $u \in V$, $|R|$ various colourings of $G$ according to $h$ exist.

$(\Leftarrow)$ Suppose that there is a function $h : T \to \{0, 1\}$ such that $k$ various colourings of the graph $G = (V, T)$, where $V = \{0, 1, \ldots, m - 1\}$, according to $h$ exist. We intend to find a set $R$, where $|R| = k$, of bit strings of length $m$ such that the last condition from Lemma \ref{lem.2} holds for the sets $R$ and $T$. We define this set as $R = \{c(0)c(1)\cdots c(m - 1) \ | \  c \ \textrm{is a colouring of} \ G \  \textrm{according to} \ h.\}$. The last condition from Lemma \ref{lem.2} holds because for each $\{i, j\} \in T $, $r_{i} \oplus r_{j} = h(\{i, j\})$ for every $r \in R$.
}

\begin{corollary}
Let $m > 0$ be an even integer. In order to show that there is no classical deterministic winning strategy for the matching game for $m$, it suffices to show that there are no graph $G = (V, E)$, where $|G| = m$, and no function $h : E \to \{0, 1\}$ such that the following conditions hold simultaneously.
\begin{enumerate}
 \item There are at least $\frac{2^{m}}{2^{\lceil log \ m \rceil}}$ colourings of $G$ according to $h$,
 \item $G$ contains a component of cardinality greater than $\frac{m}{2}$. 
\end{enumerate}
\end{corollary}

In the rest of this section, the following simple statement will be useful.

\begin{lemma}
Let $G = (V, E)$ be a graph and let $h : E \to \{0, 1\}$ be a function. If $G$ is colourable according to $h$, then there are exactly $2^{k}$ colourings according to $h$ where $k$ is a number of components of $G$.
\end{lemma}
\proof{
Suppose that there is a colouring $c$ of the graph $G$ according to $h$. It suffices to show that for any component $C = (V', E')$ of $G$, there are exactly 2 colourings of $C$ according to $h$. Since the events of colouring mutually different components of $G$ according to $h$ are independent, this gives us the desired result.
 
If $C$ contains only one vertex, the statement holds trivially because we can assign either 0 or 1 to the only vertex of $C$. Suppose further that $C$ contains $k > 1$ vertices. By restricting the colouring $c$ to the component $C$ only, we obviously obtain a colouring of $C$ according to $h$. Let us denote this restricted colouring with $c_{r}$. It is straightforward to see that a function $c_{r}': V' \to \{0, 1\}$ defined as $c_{r}'(u) = \neg c_{r}(u)$ is also a colouring of $C$ according to $h$. Now consider any colouring $q$ of $C$ according to $h$ and any vertices $u, v \in V'$. Clearly, it holds that either $q(u) = c_{r}(u)$ or $q(u) = c'_{r}(u)$. Suppose without loss of generality that the first possibility has occurred. We intend to show, using induction on the distance $d_{C}(u, v)$ of the vertices $u, v$ in $C$, that also $q(v) = c_{r}(v)$. This is certainly true for $d_{C}(u, v) = 0$ because then $u = v$. Now suppose that $d_{C}(u, v) = n > 0$ and that the equality holds for each vertex $w \in V'$ such that $d_{C}(u, v) = n - 1$. There is a path $v_{0}, v_{1}, \ldots, v_{n}$ in $C$ such that $v_{0} = u$ and $v_{n} = v$. Since the equation $q(v_{n - 1}) \oplus q(v) = h(\{u, v\})$ has to be satisfied, it follows with the help of the induction hypothesis that $$q(v) = h\{u, v\} \oplus q(v_{n - 1}) = h\{u, v\} \oplus  c_{r}(v_{n - 1}) = c_{r}(v).$$ We have shown that if the colouring $q$ agrees with the colouring $c$ on some vertex from $C$, then the two colourings agree on each vertex from $C$. A similar result can be obtained for the case of $q(u) = c'_{r}(u)$.  Consequently, we can conclude that either $d = c_{r}$ or $d = c'_{r}$.  
}

\begin{theorem}
Let $m \ge 8$ be an even integer. There are no graph $G = (V, E)$, where $|G| = m$, and no function $h : E \to \{0, 1\}$ such that the following conditions hold simultaneously.
\begin{enumerate}
 \item There are at least $\frac{2^{m}}{2^{\lceil log \ m \rceil}}$ colourings of $G$ according to $h$,
 \item $G$ contains a component of cardinality greater than $\frac{m}{2}$. 
\end{enumerate}
\end{theorem}
\proof{
Let $G = (V, E)$, where $|G| = m$, be a graph and let $h : E \to \{0, 1\}$ be a function. Suppose that there are at least $\frac{2^{m}}{2^{\lceil log \ m \rceil}}$ colourings of $G$ according to $h$. We show that the other condition cannot hold.

From the previous lemma we can conclude that the graph $G$ is composed at least of $m - \lceil log \ m \rceil$ components. Since $G$ contains a component of cardinality greater than $\frac{m}{2}$, it contains at most $\frac{m - 2}{2}$ components composed of a single vertex. This indicates that $G$ is composed at most of $\frac{m}{2}$ components. It is easy to verify that for $m \ge 8$, $\frac{m}{2} < m - \lceil log \ m \rceil$. Therefore, if $m \ge 8$, the graph $G$ cannot exist.  
}
\section{Conclusions and Open Problems}
In the present text, we have described a general model for pseudo-telepathy games and a pseudo-telepathy game called the matching game. We have dealt with the problem what the smallest size of inputs, denoted as $m$, is for which the matching game is pseudo-telepathic. We have found classical winning strategies for $m = 4$ and $m = 6$. Also, we have shown that there is no classical winning strategy for $m \ge 8$.

Since the matching game is the youngest pseudo-telepathy game, it is known very little about it so far. For example, we still do not know any nontrivial upper bound for the success of the best possible classical strategy for $m \ge 8$. This is of importance because due to erroneous measurements, it is unavoidable that Alice and Bob will not be perfect in real experiments. If they try to show that the physical world is not local realistic, it will have to be sufficient that they are significantly better than classical players could ever be. Obviously, the better Alice and Bob are than classical players, the more convincing the experiment is. 

\nonumsection{Acknowledgements}
This work has been supported by the research projects MSM0021622419 and GA\v CR 201/07/0603.

\nonumsection{References}

\end{document}